# Review Mining for Feature Based Opinion Summarization and Visualization

Ahmad Kamal
Department of Mathematics
Jamia Millia Islamia
New Delhi-110025, India
ah.kamal786@gmail.com

## ABSTRACT
The application and usage of opinion mining, especially for business intelligence, product recommendation, targeted marketing etc. have fascinated many research attentions around the globe. Various research efforts attempted to mine opinions from customer reviews at different levels of granularity, including word-, sentence-, and document-level. However, development of a fully automatic opinion mining and sentiment analysis system is still elusive. Though the development of opinion mining and sentiment analysis systems are getting momentum, most of them attempt to perform document-level sentiment analysis, classifying a review document as *positive*, *negative*, or *neutral*. Such document-level opinion mining approaches fail to provide insight about users' sentiment on individual features of a product or service. Therefore, it seems to be a great help for both customers and manufacturers, if the reviews could be processed at a finer-grained level and presented in a summarized form through some visual means, highlighting individual features of a product and users sentiment expressed over them. In this paper, the design of a unified opinion mining and sentiment analysis framework is presented at the intersection of both machine learning and natural language processing approaches. Also, design of a novel feature-level review summarization scheme is proposed to visualize mined features, opinions and their polarity values in a comprehendible way.

## General Terms
Summarization and Visualization.

## 1. INTRODUCTION
The emergence of Web 2.0 has caused rapid proliferation of e-commerce and social media contents. Web is widely used as a platform by users and manufactures to share experiences and opinions regarding products, services, marketing campaigns, social events etc. Over the past decade, the discipline of opinion mining (aka review mining) is emerging, that computationally evaluates user's opinions, subjectivity, sentiments, appraisals, emotions, feedbacks etc. expressed in customer reviews [1]. Enormous availability of customer reviews on merchant sites have attracted researchers to retrieve information for developing practical real life applications including business intelligence, product recommendation, targeted marketing, etc. Since users' opinions are very informative in developing marketing and product development plans, large business houses and corporate are taking interest in opinion mining systems. Such systems can process users' opinions and sentiments to predict better recommendations for target marketing of products and services. However, due to heterogeneity and lack of structure in customer reviews, automated distillation of knowledge is technically challenging task and requires research at the intersection of various techniques such as natural language processing, information extraction, information retrieval, data mining, machine learning etc.

In this paper, the design of a unified opinion mining and sentiment analysis framework is proposed that facilitates subjectivity/objectivity analysis, feature and opinion extraction, anaphora resolution for feature-opinion binding, polarity determination, review summarization and visualization in an integrated manner. In the first phase of the proposed approach, supervised machine learning technique is applied for subjectivity and objectivity classification of review sentences, as distillation of objective sentences improve mining performance by preventing noisy and irrelevant extraction [2]. Thereafter, natural language processing techniques are applied, covering subjective sentences of customer reviews to mine information components, which can be described as a triplet of the form <*f, m, o*>, where *f* represents a product feature, *o* represents an opinion expressed over *f*, and *m* is an optional modifier used to model the degree of expressiveness of opinion *o* [3, 28]. It has been observed that various opinions are left unnoticed due to lack of co-occurrence of features and opinions at sentence level. This occurs, when the features mentioned in a sentence are referenced in succeeding sentences using anaphoric pronouns. In order to identify the associations of such features with correct opinions, a backtracking-based anaphora resolution approach is presented for correct binding of feature-opinion pairs. Further, word-level sentiment classification scheme is exercised with the aid of statistical approach and supervised machine learning technique to determine the polarity values (*negative*, *positive* and *neutral*) of opinionated words [4]. Extracted information components with polarity values are stored in a structured format for review summary generation and visualization. Also, one of the crucial requirements when developing an opinion mining system is the ability to browse through the customer review collection and to visualize various information components present within the collection in a summarized form. A visualization technique is proposed that facilitates both customers and manufacturers in easy navigation through the pile of review documents and mining results using graphical user interfaces.

The rest of the paper is organized as follows. Brief reviews of related works conducted in this area are surveyed in Section 2. Section 3 presents the functional detail of the proposed review mining system. Feature based opinion summarization & visualization scheme is discussed in section 4. Finally, section 5 concludes the paper.

## 2. RELATED WORKS

Review mining refers to the process of extracting product features and opinions from subjective contents, computationally evaluates user's opinions, sentiments, feedbacks etc. and summarizing them using a visual representation. In the beginning of this process, subjectivity and objectivity classification is performed to distinguish between factual and subjective remarks present in customer reviews. Star-rated (1 to 5 stars) customer reviews at merchant sites help in dividing subjective and objective review documents [6]. Higher star rated documents can be placed in subjective category, whereas lower star rated documents can be assigned to objective category. However, a subjective document may also include some factual contents [5]. Thus, for better information component extraction, sentence-level subjectivity/objectivity analysis is proposed in many literatures [2, 7]. Further, extracted subjective sentences are analyzed syntactically and semantically by exploiting Parts-Of-Speech (POS) information and dependency relationship between words [8, 9, 10]. For example, product features are generally nouns, opinions are adjectives. Thus POS information based rules can be framed to analyze opinionated texts for candidate feature and opinion extraction, followed by the application of some statistical measures to identify feasible ones and discard noises [3]. It has been observed that, many features appearing as noun phrases in review sentences are generally referenced by anaphoric pronouns present in succeeding sentences of a review document [4]. As mentioned in [11], the extraction of anaphoric opinion targets has been identified as an open issue in opinion mining research, but not much research efforts have been applied in this regard. A study in [12] reported that 14% of the opinion targets (product features) are pronouns in their dataset. Thus, anaphora resolution is important for binding feature-opinion pairs, otherwise large number of opinion information will be left unnoticed. In addition to information component extraction, review mining research requires sentiment classification of every opinion bearing word present as a part of information component. In [6], unigram model is proposed using supervised learning technique for sentiment classification. However, dictionary-based [13, 14] and corpus-based [15] approaches are widely used for this purpose. Some researches present a good mix of statistical text classification methods and machine learning approaches to develop word-level sentiment classification system [4, 16].

The vast amount of opinion information available on the Web becomes astounding and creating problem for end users to browse through large collection, urging the need to visualize various information components present within the collection in a summarized form. Consequently, many researchers have used different techniques for opinion summarization and visualization. In [8] statistical summarization is adopted to represent the result of opinion mining task. Opinion summarization by tracking over a timeline is proposed in [17, 18]. Since users' opinions vary with respect to time, analyzing trends of opinions over a timeline helps in predicting users' behavior in future regarding products or services. In [19], authors divided the task of opinion summarization as single document based, multi-documents based, textual, and visual approaches. In [20], aspect-rated summary is proposed which provides a decomposed view of the overall ratings related to the major aspects (features) of a product.

For visualization purpose in [21], Gamon *et al*. adopted box and colour scheme for general assessment of product features. Shaded boxes are used to represent product features, where size of a box reflects the number of occurrences of the feature word in the underlying corpus. The color (red for negative, white for neutral, and green for positive) for any given box is used to reflect average sentiment related to the corresponding feature. Despite the large number of products under evaluation, such graphical visualization is very helpful for users in observing their positive and negative aspects. In [22], design of a prototype system *opinion observer* is presented that represents strengths and weaknesses of various product features and enables users to compare opinion information using a bar. The portion of the bar above and below a horizontal line represents the summary of statistics obtained from opinion analysis task. The graphical interface enables users to access sentiment statistics of various products in a single glance of visualization. In [23], development of *Xopin* (a graphical user interface for feature-based opinion mining system) is presented. The system allows users to browse, navigate, filter, and visualize the results of the feature-based opinion detection system. The comparison view of *Xopin* allows users to compare product features from large collection of texts vary easily.

## 3. PROPOSED REVIEW MINING SYSTEM

This section presents the architecture and functional detail of the proposed review mining system. Figure 1 presents the complete architecture of the proposed system, which consists of various functional components such as *subjectivity/objectivity analyzer*, *feature and opinion learner* that includes *rule based approach for feature-opinion pair extraction and anaphora resolution for feature-opinion binding*, *feasibility analyzer, sentiment analyzer, feature based review summarization and visualization*. Further details about these modules are presented in the following sub-sections.

## 3.1 SUBJECTIVITY & OBJECTIVITY ANALYZER

Various researches reveal that customer reviews may contain both subjective and objective contents. Subjective contents represent users' opinion, emotion, feedback, sentiment etc. whereas, objective texts reflects factual information. Thus, the target of subjectivity/objectivity classifications is to restrict unwanted and unnecessary objective texts from further processing [2]. For this purpose, each review sentence is tokenized into unigrams. Thereafter, a supervised binary classification model is implemented for classifying each word of a review sentence as subjective or objective, and consequently possibility of the enclosing sentence to be either subjective or objective is computed using a unigram model. A set of statistical and linguistic features is determined to represent unigrams as feature vectors and to learn classification models. In order to establish the efficacy of the identified features for subjectivity determination, various prominent classifiers are practiced such as Naive Bayes (a simple probabilistic classifier based on Bayes theorem) [24], J48 (a decision tree based classifier) [25], Multilayer Perceptron – MLP (a feed forward artificial neural network model with one input layer, one output layer and one or more hidden layers) [26], & Bagging (a bootstrap ensemble method) [27] implemented in WEKA [33] and 10-fold cross-validation is used for evaluation. Further details about proposed subjectivity/objectivity analysis can be found in [2].

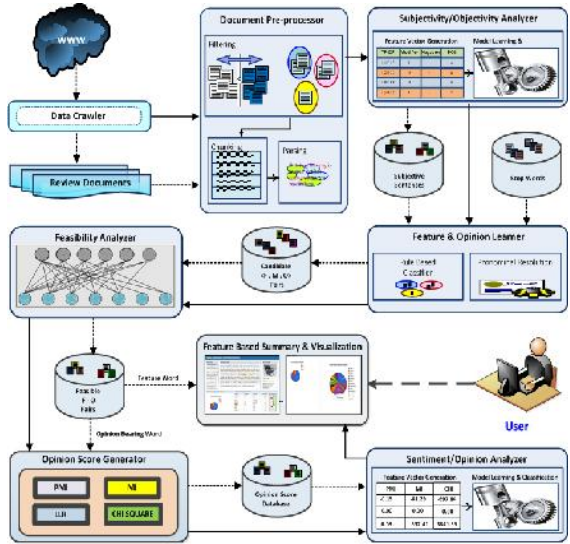

**Figure 1**: Architecture of the proposed review mining system

## 3.2 FEATURE & OPINION LEARNER

This module is responsible for the development of various approaches for information component (comprising *feature*, *modifier*, and *opinion*) extraction from subjective review sentences extracted in the previous step. In one approach, the mining process is initiated with the aid of a statistical parser and facilitated by a rule-based system to identify candidate information components for further analysis. Further, various opinions information are left unnoticed due to lack of co-occurrence of *feature-opinion* pair at sentence level. Therefore, in another approach, a backtracking technique is presented for inferring anaphora pronoun with relevant antecedent that exists in preceding sentences for binding *feature-opinion* pairs. Both approaches are generic in the sense that they can be applied on review sentences pertaining to any product domain and no prior information is needed for identification of product features and users' opinions expressed over them. Further details regarding proposed approaches can be found in [2, 3, 4].

## 3.3 FEASIBILITY ANALYZER

During information component extraction phase, various nouns, verbs and adjectives are extracted that are not relevant product features, modifiers, and opinions. Also, anaphora-antecedent binding caused noisy feature-opinion pairs extractions that are not relevant for feature-opinion binding task. In line with [3], feasibility analysis technique is applied to eliminate noisy feature-opinion pairs based on reliability scores generated through a customized version HITS algorithm [29], which models feature-opinion pairs and review document as a bipartite graph considering feature-opinion pairs as hubs and review documents as authorities. A higher score value of a pair reflects a tight integrity of the two components in a pair. In other words, this score determines the degree of reliability of an opinion expressed over a product feature. Table 1 present's hub and reliability scores for some randomly selected *feature-opinion* pairs from different electronic products.

**Table 1:** Exemplar feature-opinion pairs with hub and reliability scores

| Product Type | Feature | Opinion | Final HS (After Convergence) | Reliability Score |
|---|---|---|---|---|
| Digital Camera | Camera | Great | 18.11 | 1.00 |
| | Picture | Beautiful | 7.16 | 0.39 |
| | Photo | Good | 7.76 | 0.43 |
| | Lens | Great | 6.30 | 0.35 |
| | Video | Good | 5.78 | 0.32 |
| IPod | Ipod Apps | Coolest | 8.22 | 1.00 |
| | Camera | Great | 5.70 | 0.68 |
| | Video | Decent | 4.91 | 0.59 |
| | Sound | Richer | 1.40 | 0.17 |
| | Battery | Faulty | 1.35 | 0.16 |
| Laptop | Megapixel | Standard | 10.59 | 1.00 |
| | OS | Great | 8.48 | 0.80 |
| | Screen | Wonderful | 3.43 | 0.32 |
| | Keyboard | Great | 3.27 | 0.31 |
| | Price | Issue | 2.82 | 0.27 |
| Cell Phone | Phone | Thin | 5.70 | 1.00 |
| | OS | Tricky | 2.25 | 0.39 |
| | Screen | Large | 1.96 | 0.34 |
| | Camera | Good | 1.42 | 0.25 |
| | Keyboard | Awesome | 1.07 | 0.19 |

## 3.4 SENTIMENT ANALYZER

In addition to the extraction of *feature-opinion* pairs from review documents, another important task related to the development of an effective opinion mining system is to classify sentiment or polarity (*positive*, *negative*, or *neutral*) of opinion bearing words present as a part of information components. A supervised machine learning approach based on statistical and linguistic features for word-level sentiment classification is applied to determine the sentiments of opinionated words retained after feasibility analysis. A rich set of statistical features are identified that includes *Pointwise Mutual Information* [30], *Mutual Information* [31], *Chi-square* (commonly known as *Karl Pearson's chi-square*), and *Log Likelihood Ratio* [32]. In addition, some linguistic features are also considered, including *negation*, *tf-idf* and *modifier* for classification purpose. Table 2, shows a partial list of opinionated words and their respective opinion scores calculated using some of the statistical association functions discussed above. The proposed sentiment analyzer system is implemented as a two phase process – *model learning (aka training phase)* and *classification* (*aka testing phase*). The training phase uses the feature vectors generated from training dataset to learn the classification models, which is later used to determine the polarity of the opinionated words extracted from testing dataset. Four different classifiers are considered including Naive Bayes [24], Decision Tree (J48) [25], Multilayer Perceptron (MLP) [26] and Bagging [27], but finally settled with Decision Tree (J48) and Bagging algorithms implemented in WEKA [33] due to their best performance. Once the semantic orientation of individual opinionated words is determined, the semantic orientation can be determined at higher levels of abstraction. Further details regarding the proposed system can be found in [4].

**Table 2:** A partial list of opinionated words and their opinion scores obtained using different statistical measures

| Opinionated Word | Opinion Score | | |
|---|---|---|---|
| | PMI | MI | CHI-Square |
| Bad | -0.7344 | -109.3725 | -850.0066 |
| Expensive | -0.2984 | -51.8493 | -556.9257 |
| Poor | -0.5560 | -54.0277 | -378.8968 |
| Slow | -0.6935 | -66.1516 | -369.4841 |
| Horrible | -0.9389 | -34.7363 | -240.8866 |
| Bittersweet | 0.0000 | 0.0000 | 0.0000 |
| Unbelievable | 0.0000 | 0.0000 | 0.0000 |
| Amazing | 2.3403 | 172.7484 | 1177.2141 |
| Bright | 0.3932 | 47.6693 | 245.7392 |
| Beautiful | 1.0260 | 76.5412 | 485.5965 |
| Fantastic | 1.4603 | 98.6912 | 607.5986 |
| Wonderful | 2.0459 | 75.8369 | 419.2278 |

## 4 FEATURE BASED REVIEW SUMMRIZATION & VISUALIZATION

One of the crucial requirements when developing a review mining system is the ability to browse through the customer review collection and be able to visualize various information components present within the collection in a summarized form. Keeping in mind the above fact, the design of an *Opinion Summarization and Visualization System (OSVS)* is proposed to present extracted information components in a graphical form that facilitates users to have a quick view of a product features and users' sentiments expressed over them, without reading the pile of review documents. *OSVS* is capable to visualize mining results both from single as well as multiple review documents. It provides a graphical environment for end users to explore and visualize summarized sentiments using bar and pie charts for every product feature. *OSVS* uses Google chart API[1] to generate *bar* and *pie* charts for visualization. Extracted information components along with opinion summary statistics are presented using Java Script Object Notation (JSON)[2] object. JSON is a language independent, lightweight text-data interchangeable format. Figure 2 shows the JSON representation of an object describing information component and opinion summary statistics. The object uses string field for *feature*, *modifier, opinion* and *orientation*, a number field for *reliability score,* and contains an array of objects for *opinion score*. During execution, *OSVS* retrieves all required information from database to form JSON object and using the same as an input for visualization purpose. Figure 3 shows the main screen of *OSVS*, consisting of two rows viz. the *upper-row* and the *lower-row*. The *upper-row* is divided into three panels - *upper-left*, *upper-middle*, and *upper-right*. The *upper-left* panel contains list of reviews crawled from merchant sites. When a user selects a particular review from the *upper-left* panel, its description and metadata appears in the *upper-middle* and *upper-right* panels, respectively. Metadata of a review consists of information such as source from where the review was crawled, domain, author name, description, date of posting, and star rating. The *lower-row* of the main screen is also divided into two panels' viz. *lower-left* and *lower-right*. The *lower-left* panel uses *pie chart* for opinion summarization of a particular review selected by the end user from *upper-left* panel. The *pie chart* makes use of different colour combination mainly *blue*, *red*, and *green* to visualize

---
[1] http://code.google.com/apis/ajax/playground/#chart_wrapper
[2] http://www.json.org/

the number of *positive*, *negative*, and *neutral* opinions present in the review, respectively.

```
{
  "feature": "Speaker quality",
  "modifier": "very",
  "opinion": "bad",
  "scoreReliabilityPair": 0.0108,
  "scoreOpinion": [
    {
      "type": "pmi",
      "number": -0.7344
    },
    {
      "type": "mi",
      "number": -109.3725
    },
    {
      "type": "chi",
      "number": -850.0066
    }
  ],
  "orientation": "negative"
}
```

**Figure 2:** JSON representation of information component and opinion score

For a selected review document, the *lower-right* panel presents the list of extracted results that includes *feature*, *modifier* (if any), *opinion*, *orientation* (*positive*, *negative*, and *neutral*) and opinion indicator of the *feature-opinion* pair.

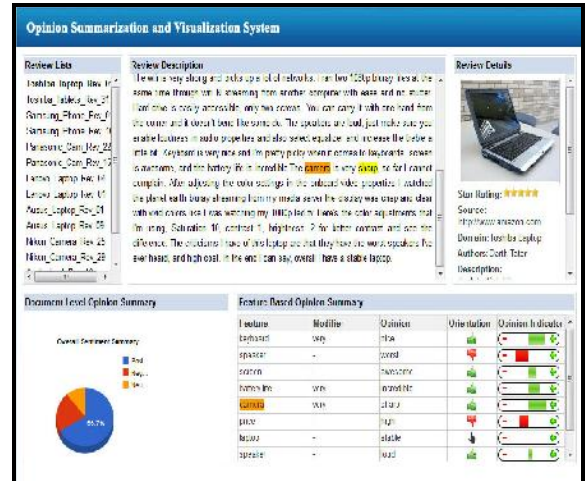

**Figure 3:** Opinion summarization and visualization using OSVS

For visibility purpose, colour scheme is used to highlight the information components extracted from review documents. On clicking a *feature* word appearing in the *lower-right* panel, the constituents of the corresponding *feature-opining* pair is highlighted using *orange* and *yellow* colors, respectively, and the relevant snippets (containing *feature* and *opinion* words) of the review document is also highlighted. When a user clicks to a highlighted snippet representing product feature in *upper-middle* panel, a pop-up window appears visualizing the percentage of *positive*, *negative*, and *neutral* opinions using *pie-chart*. Figure 4 shows the percentage of opinions expressed on a product feature from a corpus of customer reviews. As discuss earlier, *OSVS* facilitates users to navigate through the pile of customer reviews in an efficient way to produce feature-based opinion summary.

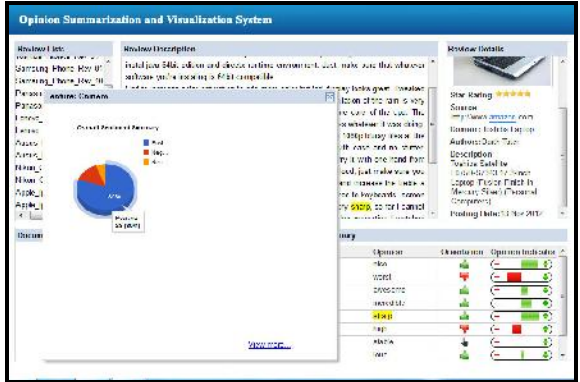

**Figure 4:** Feature-based opinion summarization using OSVS

Thus, pop-up window appearing in the above mentioned step contains a *view more* option, clicking which causes the window to expand in size, and visualizing opinion score summary for the respective product feature. Figure 5 shows an expanded pop-up window, where size of each slice in the *3D pie-chart* represents the degree of expressiveness of opinion. Opinion scores are calculated using *Chi-square* value due to its best performance. Higher the opinions score for an opinion bearing word, larger the size of a slice in the *3D pie-chart*.

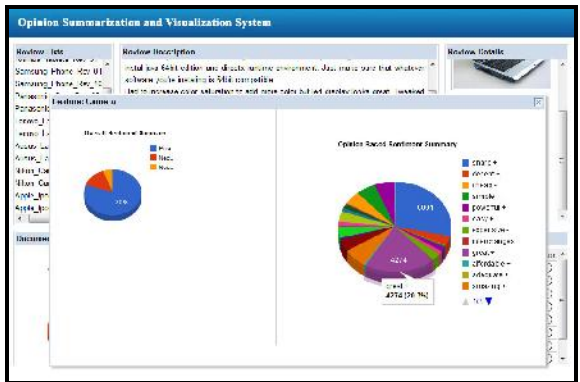

**Figure 5:** Opinion based sentiment summary using OSVS

## 5. CONCLUSION

Since last decade, the application and usage of opinion mining have fascinated many research attentions around the globe. Various research efforts attempted to mine opinions from customer reviews. However, development of a fully automatic opinion mining and sentiment analysis system is still elusive. Rapid growth of unstructured or semi-structured user-generated contents on the Web and their uncontrolled generation consisting of various natural language nuances possesses a big challenge on research community in fully automating information component extraction. It has been observed that overall problems associated with opinion mining and sentiment analysis is non-trivial and requires more research exploration. The main contribution of this work remains in studying feature-based opinion mining and sentiment classification from review documents at finer-grained level and finally coming up with methods for distinguishing subjective and objectivity sentences, feature-opinion pair extraction, and sentiment classification. In this paper, the design of feature based opinion summarization and visualization system is presented to facilitate the visualization and summarization of review mining results in a graphical form. The system represents extracted information components and opinion scores as a JSON object, and uses the same as an input for visualization purpose. Various graphical entities such as *bar* and *pie* charts are employed for visualization purpose. Colour scheme is used to highlight the extracted information components from review documents. The proposed *OSVS* is capable of visualizing opinion mining results both from single as well as multiple review documents.